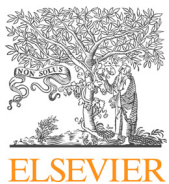

# High-performance magnesium/sodium hybrid ion battery based on sodium vanadate oxide for reversible storage of Na$^+$ and Mg$^{2+}$

Xiaoke Wang [a,b,c], Titi Li [b], Xixi Zhang [b], Yaxin Wang [c], Hongfei Li [c], Hai-Feng Li [a,*], Gang Zhao [b,*], Cuiping Han [d,e,*]

[a] Institute of Applied Physics and Materials Engineering, University of Macau, Avenida da Universidade, Taipa, Macao 999078, China
[b] School of Physics and Technology, University of Jinan, Jinan 250022, Shandong, China
[c] Songshan Lake Materials Laboratory, Dongguan 523808, Guangdong, China
[d] Faculty of Materials Science and Energy Engineering, Shenzhen University of Advanced Technology, Shenzhen 518055, Guangdong, China
[e] Institute of Technology for Carbon Neutrality, Shenzhen Institute of Advanced Technology, Chinese Academy of Sciences, Shenzhen 518055, Guangdong, China

ABSTRACT

Magnesium ion batteries (MIBs) are a potential field for the energy storage of the future but are restricted by insufficient rate capability and rapid capacity degradation. Magnesium-sodium hybrid ion batteries (MSHBs) are an effective way to address these problems. Here, we report a new type of MSHBs that use layered sodium vanadate ((Na, Mn)V$_8$O$_{20}$·5H$_2$O, Mn-NVO) cathodes coupled with an organic 3,4,9,10-perylenetetracarboxylic diimide (PTCDI) anode in Mg$^{2+}$/Na$^+$ hybrid electrolytes. During electrochemical cycling, Mg$^{2+}$ and Na$^+$ co-participate in the cathode reactions, and the introduction of Na$^+$ promotes the structural stability of the Mn-NVO cathode, as cleared by several ex-situ characterizations. Consequently, the Mn-NVO cathode presents great specific capacity (249.9 mA h g$^{-1}$ at 300 mA g$^{-1}$) and cycling (1500 cycles at 1500 mA g$^{-1}$) in the Mg$^{2+}$/Na$^+$ hybrid electrolytes. Besides, full battery displays long lifespan with 10,000 cycles at 1000 mA g$^{-1}$. The rate performance and cycling stability of MSHBs have been improved by an economical and scalable method, and the mechanism for these improvements is discussed.

© 2024 Science Press and Dalian Institute of Chemical Physics, Chinese Academy of Sciences. Published by ELSEVIER B.V. and Science Press. All rights reserved.

## 1. Introduction

Rechargeable magnesium ion batteries (MIBs) are favorable electrochemical energy storage systems that can meet future electrical energy storage requirements [1,2] due to their potential advantages, such as the large theoretical volumetric capacity of magnesium (3833 mA h cm$^{-3}$) and minimal environmental impact [2,3]. Magnesium resources are abundant, evenly distributed, easy to mine on earth, and have an inherent cost advantage [4,5]. Nevertheless, due to their small ionic radius as well as multi-electron redox reactions, metal cations like Mg$^{2+}$ exhibit a high charge density (0.72 Å for Mg$^{2+}$, 0.76 Å for Li$^+$) [6,7]. The high charge density of Mg$^{2+}$ leads to sluggish solid-state diffusion, resulting in strong polarization and low intercalation efficiency in electrode materials [8,9]. Despite considerable effort, these reversibility advances are mostly achieved at a slower rate (<0.5 C) [10]. Therefore, MIBs suffer from insufficient cycling and rate performance [11]. It remains an obvious challenge for MIBs to obtain excellent performance in practical applications [12,13].

To date, a variety of tactics have been pursued thus far to raise the cycling stability and rate capability of MIBs [14], containing interface engineering of magnesium anodes [15], electrode material structure modification [9], electrolyte regulation [2], capacity-compensation emerging from the dynamic redox of copper ions [16], and so on. A suitable solution is to add alien cations (e.g., Li$^+$, Na$^+$) for better kinetic performance. Such batteries can be called magnesium-lithium/sodium hybrid ion batteries (MLHBs, MSHBs) [17]. Previously, an efficient method to address this issue is to build MLHBs that benefit from a dendrite-free magnesium anode and accelerated Li$^+$ kinetics facilitated by a dual-salt electrolyte that contains Mg$^{2+}$ and Li$^+$ ions [18,19]. Compared with MLHBs, MSHBs have been relatively less studied and the obtained cycle stability or specific capacity in MSHBs remains unsatisfactory [20]. As shown in Table S1, a type of sulfides and oxides cathode, such as FeS$_2$ [21], Na$_2$VTi(PO$_4$)$_3$ [22], Mn$_3$O$_4$ [23], Mg$_{1.5}$VCr(PO$_4$)$_3$ [20], and FeFe(CN)$_6$ [24], has been prepared in the MSHBs. However, some of the cathode materials have limited cycle life or insufficient discharge capacity. As a result, there is a pressing demand for cathode materials that exhibit improved performance.

\* Corresponding authors.
 *E-mail addresses:* haifengli@um.edu.mo (H.-F. Li), sps_zhaog@ujn.edu.cn (G. Zhao), cp.han@siat.ac.cn (C. Han).

Herein, sodium vanadate oxide ((Na, Mn)V$_8$O$_{20}$·5H$_2$O, Mn-NVO) materials were proposed as the cathode materials for MSHBs. The Mn-NVO material has an attractive layered structure and the spaces between the metal oxide layers naturally form a two-dimensional diffusion pathway for Mg ion (de)intercalation [25]. Na ions, which possess the advantage of rapid diffusion kinetics, were further introduced through the utilization of a dual salt electrolyte. The synergistic motion between Mg$^{2+}$ and Na$^+$ ions improves the diffusion kinetics, allowing the Mn-NVO cathode to have exceptional rate capability and cycling stability. Specifically, Mn-NVO cathode demonstrates notable performance characteristics, including a reversible capacity of 249.9 mA h g$^{-1}$ at 300 mA g$^{-1}$, enhanced rate performance (130.2 mA h g$^{-1}$ at 300 mA g$^{-1}$), and durable cyclic performance (1500 cycles at 1500 mA g$^{-1}$ with 88.2% capacity retention), which represents a considerable improvement compared to the up-to-date cathode of MSHBs. Ex-situ characterizations have revealed that Mg$^{2+}$ and Na$^+$ co-participate in the cathode reactions and Mn-NVO cathode endures a reversible phase change throughout the charging and discharging processes. Lastly, the combination of Mn-NVO as the cathode and organic 3,4,9,10-perylenetetracarboxylic diimide (PTCDI) as the anode in a full MSHB achieves a remarkable cycle stability of 10,000 cycles at 1000 mA g$^{-1}$. This work presents a novel excellent performance and high-safe MSHB system that has a large potential for future energy storage applications.

## 2. Experimental

### 2.1. Sample preparation

The synthesis of Mn-doped NaV$_8$O$_{20}$·5H$_2$O (Mn-NVO) nanobelts was performed using a one-step hydrothermal method. Initially, a mixture of 4.73 g of V$_2$O$_5$ (Aladdin, AR, 99%), 1.99 g of Na$_2$SO$_4$ (Aladdin, AR, 99%), and 0.676 g of MnSO$_4$·H$_2$O (Sinopharma Co., Ltd. Shanghai, China, AR) was dispersed in 50 mL of deionized water. Subsequently, the pH of the solution was adjusted to 3 by adding CH$_3$OOH (Aladdin, ≥99.8%) while stirring in a magnetic stirrer for 30 min. The resulting solution was then transferred to a 100 mL Teflon-lined stainless-steel autoclave and heated at 180 °C for 72 h. Finally, the Mn-NVO samples were dried overnight at 60 °C in a vacuum oven.

### 2.2. Structural characterizations

X-ray diffraction (XRD) analysis was executed using a D8 Advance X-ray diffractometer (MiniFlex600 with Cu $K_\alpha$ radiation) with an angle range from 3° to 90°. X-ray photoelectron spectroscopy (XPS) characterizations were executed using the PHI 5000 VersaProbe II spectrometer equipped with Al $K_\alpha$ radiation. Scanning electron microscopy (SEM) was conducted by scanning electron microscopy (SEM, HITACH S4800, 10 kV, 5 mA). High-resolution transmission electron microscopy (HRTEM) and transmission electron microscopy/energy dispersive spectroscopy (TEM-EDS) mapping were conducted using the FEI Tecnai G2 F30 transmission electron microscope (200 kV).

### 2.3. Electrochemical characterizations

In order to fabricate the cathode, a mixture of the prepared Mn-NVO, Ketjen black, and polyvinylidene fluoride (PVDF) binder was prepared in a weight ratio of 7:2:1. The components were mixed together in N-methyl-2-pyrrolidone (NMP) solvent to form a slurry. This slurry was then coated onto a current collector made of carbon cloth (Ce Tech Co., Ltd). The fabricated electrodes were subsequently dried overnight in a vacuum oven at 60 °C. The electrodes were loaded with active materials at a mass loading of 1–2 mg cm$^{-2}$. In order to fabricate the anode, a mixture of PTCDI, Ketjen black, and PVDF binder was prepared in a weight ratio of 7:2:1. The components were combined in NMP solvent to form a slurry. This slurry was then applied onto a carbon cloth current collector obtained from carbon cloth (Ce Tech Co., Ltd). The resulting electrodes were dried overnight in a vacuum oven at 60 °C. The active materials were loaded onto the electrodes with a mass loading of 4–5 mg cm$^{-2}$.

The performance of the prepared Mn-NVO cathode was evaluated electrochemically using three-electrode cells. The Mn-NVO cathode served as the working electrode, a platinum electrode functioned as the counter electrode, and an Ag/AgCl electrode was utilized as the reference electrode. The electrolyte solution was formulated by dissolving 1 M Mg(ClO$_4$)$_2$·6H$_2$O and 1.5 M NaClO$_4$ in a mixed solvent comprising tetraethylene glycol dimethyl ether (TEGDME) and water, with varying volume ratios (TEGDME: H$_2$O = 0.8: 0.2). The 1 M Mg(ClO$_4$)$_2$·6H$_2$O dissolved in TEGDME: H$_2$O = 0.8: 0.2, 1 M Mg(ClO$_4$)$_2$·6H$_2$O + 0.5 M NaClO$_4$ dissolved in TEGDME: H$_2$O = 0.8: 0.2, 1 M Mg(ClO$_4$)$_2$·6H$_2$O + 1 M NaClO$_4$ dissolved in TEGDME: H$_2$O = 0.8:0.2, and 1.5 M NaClO$_4$ dissolved in TEGDME: H$_2$O = 0.8: 0.2 were uses as the controlling electrolyte. Following that, CR-2032 coin cells were fabricated with the Mn-NVO as the cathode! PTCDI as the anode, and 100 μL of 1.5 M NaClO$_4$ and 1 M Mg(ClO$_4$)$_2$ in TEGDME: H$_2$O = 0.8: 0.2 as electrolytes.

Cyclic voltammetry (CV) tests of the Mn-NVO cathode, PTCDI anode, and the Mn-NVO//PTCDI full cell were performed using an electrochemical station (CHI760E). The scan rates of Mn-NVO cathode ranged from 0.5 to 5 mV s$^{-1}$, and the voltage range was −0.7 to 1.2 V vs. Ag/AgCl. The scan rates of PTCDI anode employed were 0.1, 0.3, 0.5, 0.8, and 1.0 mV s$^{-1}$, while the voltage range spanned from −1.1 to 0.2 V vs. Ag/AgCl. The scan rates of the Mn-NVO//PTCDI full battery employed were 1.0, 2.0, 3.0, 4.0, and 5.0 mV s$^{-1}$, while the voltage range spanned from 0 to 1.8 V. The Mn-NVO cathode, PTCDI anode, and Mn-NVO//PTCDI full battery underwent the galvanostatic charge and discharge experiments using a land battery tester (CT2001A). The Mn-NVO cathode at various current densities ranged from 0.3 to 5.0 A g$^{-1}$. The voltage range for the tests was −0.7 to 1.2 V vs. Ag/AgCl. The current density of the PTCDI anode ranged from 0.3 to 5.0 A g$^{-1}$, and the voltage range for the tests was −1.1 to 0.2 V vs. Ag/AgCl. The current density of the Mn-NVO//PTCDI full battery ranged from 0.3 to 5.0 A g$^{-1}$, and the voltage range for the tests was 0 to 1.75 V. The impedance spectra were measured at Gamry with a frequency sweep range of 0.01–100,000 Hz.

The electrochemical process of the (Na, Mn)V$_8$O$_{20}$·5H$_2$O electrode material involves eight electrons. We have recalculated the theoretical capacity based on this understanding, and it amounts to 239.3 mA h g$^{-1}$. F = Faraday constant (96485 C mol$^{-1}$), M = Molar mass (g mol$^{-1}$), n = Electron transfer number.

$$\text{Theoretical capacity } \left(\text{mA h g}^{-1}\right) = \frac{F \times n}{3.6 \times M} \tag{1}$$

The thermodynamic voltage-composition relationship referring to the equilibrium phase diagram of the system was found using the galvanostatic intermittent titration technique (GITT). The GITT data were used to compute the interfacial reaction resistances ($R_s$, cm$^{-2}$) of the Mn-NVO cathode during the charge/discharge operation. GITT tests were conducted under the current density ($I$) of 0.3 mA cm$^{-2}$ to charge and discharge. First, the battery was dis-charged (charged) for 5 min, and then rested until the battery voltage was stabilized (∼10 min). Second, it continued to discharging (charging) for 5 min, and repeated the above process until the cut-off voltage was −0.7–1.2 V (vs. Ag/AgCl). The reaction resistance can be calculated by the instantaneous voltage drop (ΔE) and current density (I) of discharge (charge) [26].

$$\Delta E = E_0 - E_1 = I \times R_s \tag{2}$$

$$R_S = \frac{E_0 - E_1}{I} \tag{3}$$

## 3. Results and discussion

The transportation of $Mg^{2+}$ ions through electrode materials can be challenging due to stronger electrostatic interactions. This can lead to issues such as strong polarization, electrolyte decomposition, irreversible phase transformations, or difficulties in intercalation within electrode materials during discharge-charge processes [27]. Therefore, we took advantages of the MSHBs, which used a $Mg^{2+}/Na^+$ hybrid electrolyte to enhance the kinetics of the $Mg^{2+}$ ions, achieving excellent rate and cycling stability (Fig. 1) [28]. MSHBs are more advantageous in terms of economy and safety and are well suited for wide storage applications [29]. To construct MSHBs, the rapid diffusion kinetics of Na ions was harnessed by incorporating them into MIBs through a dual salt electrolyte [30]. Besides, in a "rocking chair" type battery, both metal cations can been used as carriers and take part in the half-cell reaction by the hybrid electrolyte [6]. The hybrid ion systems may improve rate performance due to the synergistic motion between ions that promote kinetics [6,31]. The MSHBs leverage the faster diffusion rate of $Na^+$ in solids to improve the slow diffusion kinetics, enhance the rate performance in MIBs, and maintain a relatively high specific capacity.

Layered vanadium oxide has demonstrated significant potential as a cathode material in magnesium ion batteries and was therefore chosen as the model material for this study [25,32]. The XRD pattern of the Mn-NVO was demonstrated in Fig. 2(a), revealing a total of eight diffraction peaks observed at 2θ values of 8.2°, 24.7°, 25.6°, 33.1°, 34.6°, 41.8°, 50°, and 61.3°, corresponding to the reflection planes of (0 0 1), (0 0 3), (1 1 0), (0 0 4), (1 1 −3), (0 0 5), (0 2 0), and (0 2 4), consistent with the (Na, Ca)(V, Fe)$_8$O$_{20}$·$n$H$_2$O belonging to the C2/m space group (JCPDS No. 45-1363) [33]. The well-defined and intense diffraction peaks indicate highly crystalline nature of Mn-NVO [23].

The morphology of Mn-NVO was explored using SEM and TEM measurements (Fig. 2a and b). As depicted in Fig. 2(a), the Mn-NVO particles appear as uniform nanobelts. The SEM image in Fig. 2(a) reveals that the Mn-NVO particles consist of nanobelts with a diameter of ranging from 150 to 250 nm, which agree with the TEM images (Fig. 2b). Besides, the HRTEM image in Fig. 2(b) reveals the remarkably crystalline structure of Mn-NVO nanobelts. The crystal lattice in the image reveals interlayer distance of 0.21 and 0.15 nm, matching the (0 0 5) and (7 1 0) planes of NVO, respectively [22]. The TEM-EDS images depicted in Fig. S1 reveal the homogeneous distributions of V, O, Na, and Mn within the Mn-NVO. Mn-NVO has a typical layered crystal structure; the introduction of Mn ion results in the displacement of Na (1) from its average position; and the position previously occupied by Na (1) is now taken over by Mn (1), demonstrating that Mn-NVO adopts the crystal structure of (Na, Mn)V$_8$O$_{20}$·$n$H$_2$O (Fig. 2c) [34]. The nanoscale structure of Mn-NVO is considered to be useful in shortening the ion and electron transport paths and in helping attenuate the irreversible deformation caused by the polarization phenomenon [1]. In addition, since the doping amount of Mn is very limited, its contribution to the capacity is negligible.

To determine the optimal amount of $Na^+$ in the hybrid electrolyte, different concentrations of sodium salts (0.5, 1.0, and 1.5 M NaClO$_4$) were introduced into the $Mg^{2+}$ based electrolyte (i.e., 1 M Mg(ClO$_4$)$_2$ in TEGDME: H$_2$O = 4: 1). We selected TEGDME/H$_2$O (4:1) as the solvent due to its ability to provide a larger electrochemical window and higher ionic conductivity [25,35,36]. Additionally, the electrochemical performance of the NVO cathode in different concentrations of Mg(ClO$_4$)$_2$ in the hybrid electrolyte was compared (Fig. S2). When utilizing 0.5, 1.0, and 1.5 M NaClO$_4$ in MSHBs, the discharge capacity of Mn-NVO exhibits a rise to 217.7, 238.5, and 243.6 mA h g$^{-1}$, respectively (Fig. S3). Mn-NVO presents great electrochemical activity when paired with a 1.5 M NaClO$_4$ electrolyte, so 1.5 M NaClO$_4$ in MSHBs was selected as the optimal electrolyte [10]. Besides, Fig. 2(d) presents the rate performance of the Mn-NVO cathode when tested with different electrolytes (i.e., 1 M Mg(ClO$_4$)$_2$, 1.5 M NaClO$_4$, and 1 M Mg(ClO$_4$)$_2$ + 1.5 M NaClO$_4$ in TEGDME: H$_2$O = 4: 1). Rate performance of Mn-NVO cathode in MSHBs (1 M Mg(ClO$_4$)$_2$ + 1.5 M NaClO$_4$ in TEGDME: H$_2$O = 4: 1) is better than that of the Mn-NVO in SIBs (1.5 M NaClO$_4$ in TEGDME: H$_2$O = 4: 1) and MIBs (1 M Mg(ClO$_4$)$_2$ in TEGDME: H$_2$O = 4: 1). At 300 mA g$^{-1}$, it exhibits a great capacity of 249.9 mA h g$^{-1}$. As discharge rates increase to 3000 mA g$^{-1}$, the capacity retains 130.2 mA h g$^{-1}$, implying the advantage of fast charging [37]. Besides, at 300 mA g$^{-1}$, it presents great capacity recoverability. By contrast, the Mn-NVO cathode in the SIBs exhibits a significantly lower capacity of 70.6 mA h g$^{-1}$ at 300 mA g$^{-1}$, which is not competitive for practical applications. Besides, the reversible capacities in the MIBs can only maintain ∼71.8 mA h g$^{-1}$ at 3000 mA g$^{-1}$, which is attributed to the inherently slow electrochemical kinetics. The impressive capacities for MIBs at less than 1000 mA g$^{-1}$ provide an evidence of the successful insertion of $Mg^{2+}$ ions into the Mn-NVO cathode material. Nevertheless, as the current density further increases (i.e., 1500, 2000, 2500, and 3000 mA g$^{-1}$), the specific capacity significantly decreases, showing a relatively low specific capacity. Mn-NVO cathode in MSHBs has excellent rate performance compared to that in MIBs, especially at high current densities [38]. This suggests that the Mg ions inserting into the Mn-NVO cathode interlamination becomes challenging under significant polarization [8]. Additionally, the charge/discharge curves of the Mn-NVO cathode are 61.3, 112.8, and 145.2 mA h g$^{-1}$ at 2000 mA g$^{-1}$ in the SIBs, MIBs, and MSHBs, respectively (Fig. 2e). The MSHBs exhibit excellent specific capacity at higher currents, indicating their outstanding rate performance. Notably, MSHBs display a higher voltage plateau, suggesting a synergistic effect between magnesium and sodium ions [39]. In addition, MSHBs have two voltage platforms, which further demonstrates the co-intercalation of Mg$^{2+}$ and Na$^+$ in the Mn-NVO cathode.

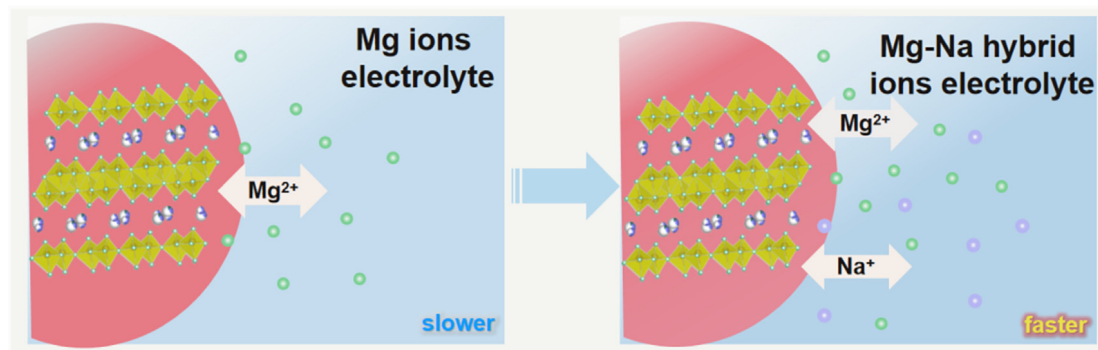

**Fig. 1.** Schematic diagram illustrating the inclusion of $Na^+$ ions in MSHBs to enhance the inadequate rate performance of MIBs.

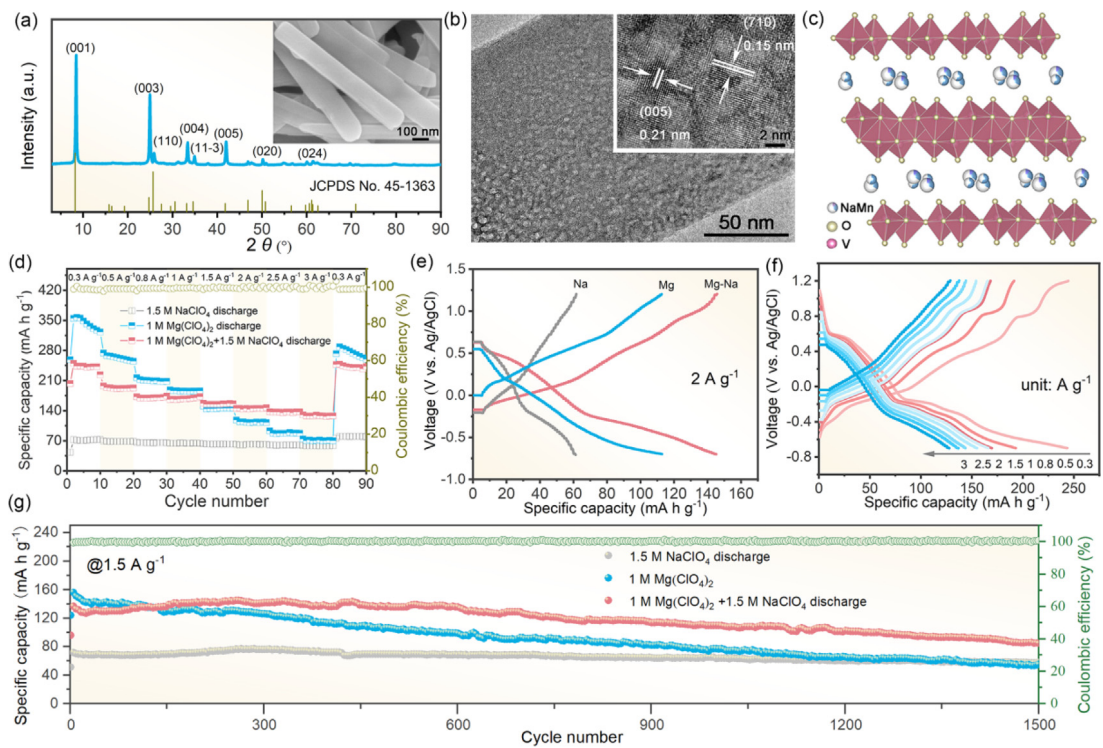

**Fig. 2.** (a) XRD diffraction pattern and SEM image of Mn-NVO; (b) TEM images and HRTEM images of Mn-NVO; (c) crystal structure of Mn-NVO; (d) the rate performances of Mn-NVO cathode in SIBs, MIBs, and MSHBs with the corresponding Coulombic efficiency in MSHBs; (e) discharge-charge curves of the Mn-NVO cathode in SIBs, MIBs, and MSHBs; (f) discharge-charge curves of the Mn-NVO cathode at various rates in MSHBs; (g) cycling performance of Mn-NVO cathode in SIBs, MIBs, and MSHBs at 1.5 A g$^{-1}$.

Fig. 2(f) displays the relevant charge/discharge profiles of the Mn-NVO cathode in MSHBs. The results show a discharge capacity of 249.9 mA h g$^{-1}$ when operated at 300 mA g$^{-1}$. Notably, even under 3000 mA g$^{-1}$, it holds a discharge capacity of 130.2 mA h g$^{-1}$. Besides, the discharge capacities for MSIBs at 0.5, 0.8, 1, 1.5, 2, and 2.5 A g$^{-1}$ are 193, 169.5, 166, 155, 143.6, and 136.8 mA h g$^{-1}$, respectively. The sloping charge-discharge plots observed indicate an ion (de)intercalation process characterized by a solid-solution type behavior [40], which is better than those in MIBs and SIBs. This reveals that there is a synergistic effect between Mg$^{2+}$ and Na$^+$ in the hybrid electrolyte. Simultaneously, after 1500 cycles at 1.5 A g$^{-1}$, Mn-NVO cathode fluctuates slightly at the initial stage (~300 cycles) and fails quickly after 1500 cycles with a capacity decaying to 54.9 mA h g$^{-1}$ in MIBs [41].

In addition, attenuation is only employed to reach capacity without impacting the well-shaped steady voltage plateau during the cycle process (Fig. 2f) [41]. At higher current densities, it is likely that the NVO cathode may not undergo sufficient reaction, leading to discrepancies in the charge/discharge curves. This phenomenon can be attributed to several factors. Firstly, the rapid insertion of ions onto the surface of the NVO active material at higher current densities may result in the deactivation of the outermost layer. Consequently, subsequent ions may not have adequate time to migrate into the depths of the active material. Additionally, the heightened currents may induce a higher degree of disruption, increasing the likelihood of damaging the crystalline structure of the NVO cathode.

Although the Mn-NVO cathode presents good cycling stability in SIBs, its insufficient capacity (57.6 mA h g$^{-1}$ after 1500 cycles) remains a challenge for widespread adoption in large-scale energy storage applications. While simple MIBs demonstrated high capacity, they faced challenges related to rate performance and cycle stability. Conversely, SIBs prepared with only sodium salt in the electrolyte showed poor capacity but exhibited excellent rate performance and cycle stability. To address these limitations, we adopted the strategy of MSHBs, combining the advantages of both systems. The cell preserved 84.6 mA h g$^{-1}$ with near 100% coulombic efficiency, exhibiting both the structural stability of the Mn-NVO cathode and the cycling stability of the MSHBs (Fig. 2g) [42]. As a result, the stability of the Mn-NVO cathode in MSHBs can be attributed to a minimized presence of free water molecules, which minimizes irreversible parasitic reactions, and the emergence of synergistic effect between Mg$^{2+}$ and Na$^+$ ions.

In order to explore the kinetic variance, CV curves were obtained for the Mn-NVO cathode in both MIBs and MSHBs at 5.0 mV s$^{-1}$ (Fig. 3a) [25]. Indeed, the discharge specific capacity of MIBs is notably higher than that of MSHBs at low currents. However, the CV measurement reveals far more distinct redox peaks for Mn-NVO cathode in MSHBs related to the reversible reaction in comparison with that in MIBs, demonstrative of significantly superior kinetics and redox reversibility in MSHBs [38]. To disclose the causes of the great rate performance of MSHBs, the electrochemical kinetics was explored through CV curves from 1.0 to 5.0 mV s$^{-1}$ (Fig. 3b). Even at higher scan rates, the CV curves of the Mn-NVO cathode still retained clear redox peaks, implying a great rate capability [38]. Besides, just a small peak shift was observed within 1.0 to 5.0 mV s$^{-1}$, demonstrating a great electrochemical reversibility [43].

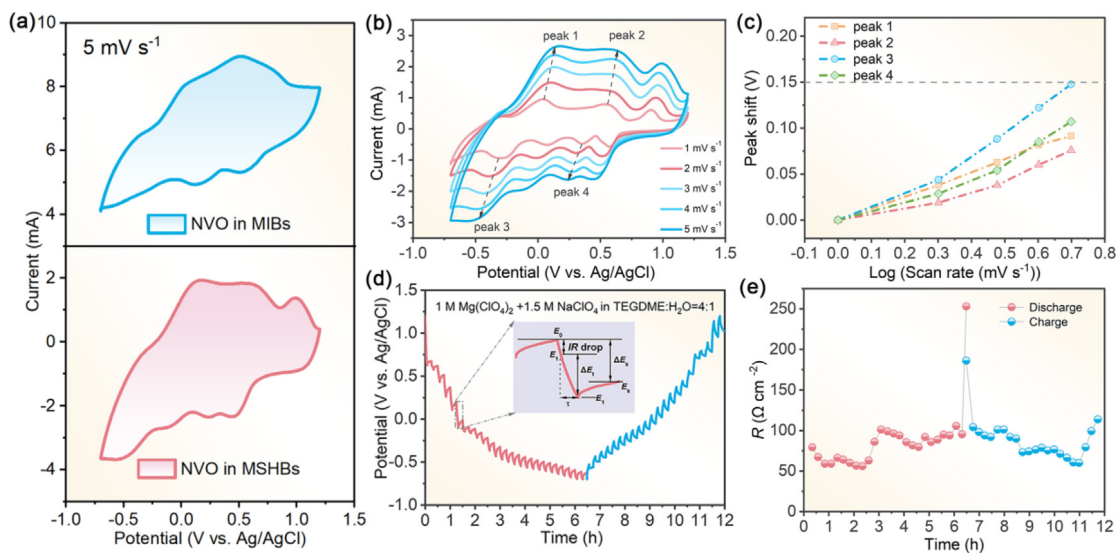

**Fig. 3.** (a) Contrast of two CV curves of the half-cells in MIBs and MSHBs at 5.0 mV s$^{-1}$; (b) CV curves of Mn-NVO at 1.0, 2.0, 3.0, 4.0, and 5.0 mV s$^{-1}$ in MSHBs; (c) cathodic peak potential shifts vs. log (scan rate) curves; (d) GITT curves and (e) changes in reaction resistance during the discharge/charge process of a half-cell in MSHBs at 0.3 mA cm$^{-2}$.

Generally, the correlation between peak currents (*i*) and the scan rates (*v*) can be mathematically represented by the following equation [25].

$$i = av^b \qquad (4)$$

The calculated *b* values in Fig. S4 are 0.67, 0.77, 0.70, and 0.82, respectively, implying a portion of the capacitive behavior reaction in MSHBs [44], which helps the fast diffusion of ions to the Mn-NVO cathode [45]. Moreover, as demonstrated in Fig. 3(c), minimal shifts in peak potentials (<0.15 V) are discovered for both cathodic and anodic peaks from 1.0 to 5.0 mV s$^{-1}$, which benefits fast energy storage [43]. The initial five CV curves of the Mn-NVO cathode achieved at 1 mV s$^{-1}$ in hybrid electrolytes with different sodium contents are revealed in Figs. S5 and S6. The CV curve of the second cycle for the Mn-NVO cathode closely resembles that of the first cycle, indicating good cycling stability. During the initial cathodic scan, the Mn-NVO cathode displayed four distinct cathode peaks at potential values of 0.94, 0.70, 0.61, and 0.37 V. Subsequent anodic scans exhibited four peaks at 0.49, 0.71, 0.92, and 1.1 V. These profiles suggest multiple phase transitions, a characteristic feature of Na$^+$ intercalation in layered materials. The presence of both reduction and oxidation peaks in pairs signifies a multi-step reaction mechanism associated with the co-insertion/extraction of Mg$^{2+}$/Na$^+$ ions, which is different from the V$_2$O$_5$ based electrodes that present two representative pairs of redox peaks [46].

The GITT was employed to conduct in situ monitoring of the interfacial reaction resistances during diverse (dis)charge stages of the Mn-NVO electrode. For the GITT measurements, a pulse duration of 5 min with a rest period of 10 min was used to collect the potential response while applying a current density of 0.3 mA cm$^{-2}$ (Fig. 3d). Fig. 3(e) shows reaction resistances derived from the (dis)charge operations. It appears that the reaction resistances experience a slight decrease during the first discharge potential (0.7–−0.2 V vs. Ag/AgCl), followed by significant raise during the final discharge potential (−0.2–−0.7 V vs. Ag/AgCl). Furthermore, this process exhibits reversibility during the charging process, confirming that the changes occurring in the Mn-NVO electrode are periodic and reversible [26].

Briefly, the addition of Na$^+$ enhances the structural stability, promoting the dynamics of the Mg$^{2+}$ reaction and improving the reversibility [47]. Moreover, the higher non-diffusion limited capacity contributions as well as outstanding dynamic behaviors allow excellent electrochemistry performance of the Mn-NVO cathode in MSHBs [43].

To better perceive the ion storage mechanism of Mn-NVO cathode in MSHBs, the investigation of the reaction mechanisms involved ex-situ XRD, the related charge-discharge curve at 0.2 A g$^{-1}$, and ex-situ XPS (Fig. 4) [48]. The ex-situ XRD patterns of MIBs and MSHBs are indicated in Fig. 4(a–d) [49]. Firstly, ex-situ XRD patterns of Mn-NVO in MIBs at the states as marked in Fig. 4(a) are exhibited in Fig. 4(b) [30]. As observed in Fig. 4(a and b), the XRD is relatively reversible in the initial cycle in MIBs as discussed in the previous work [25]. The irreversibility of XRD in the second cycle suggests that there is a need to improve the reversibility and stability of Mn-NVO in MIBs. The unfriendly reversibility and structural change during the second cycle process may be caused by two points. One reason is that the inserted Mg$^{2+}$ into the Mn-NVO cathode leads to larger phase and crystal structure changes, which destroys the original structure; the other is that the insertion of Mg$^{2+}$ can cause lattice strain in the Mn-NVO cathode, leading to distortion within its lattice structure. Both of these causes can lead to a structural collapse or a pulverization of the Mn-NVO cathode in MIBs, resulting in the poor reversibility and cycling stability [50].

Next, we adopted the strategy of MSHBs to solve the problems associated with MIBs. Fig. 4(c and d) presents the charge-discharge curves of Mn-NVO and related sites that were investigated using ex-situ XRD analysis to explore its structural change in MSHBs [30]. Firstly, the pristine material well matches the (Na, Ca)(V, Fe)$_8$-O$_{20}$·nH$_2$O with the C2/m space group (JCPDS No. 45-1363) [33], which is consistent with Mn-NVO (Fig. 4d) [34]. The initial Mn-NVO is filled with Na$^+$. The purpose of the first charging is to allow extraction of the Na$^+$ ions from the interlayer, which provides the space for the following intercalation of Mg$^{2+}$ and Na$^+$. When compared to the initial Mn-NVO cathode, a new phase appeared at 5.3°, 10.7°, and 21.1° for the first full charged material (C$_{2.7}$H$_{8.1}$O$_{1.35}$S$_{1.35}$!V$_2$O$_5$, JCPDS No. 45-1529), and it returned to its original position of 8.1° after discharge. After the subsequent fully discharged, it remained reversible in the second cycles. Therefore, throughout the charging and discharging processes, a reversible phase transition process existed, demonstrating an outstanding structural stability and reversibility of the Mn-NVO cathode with respect to the host material in MSHBs [42].

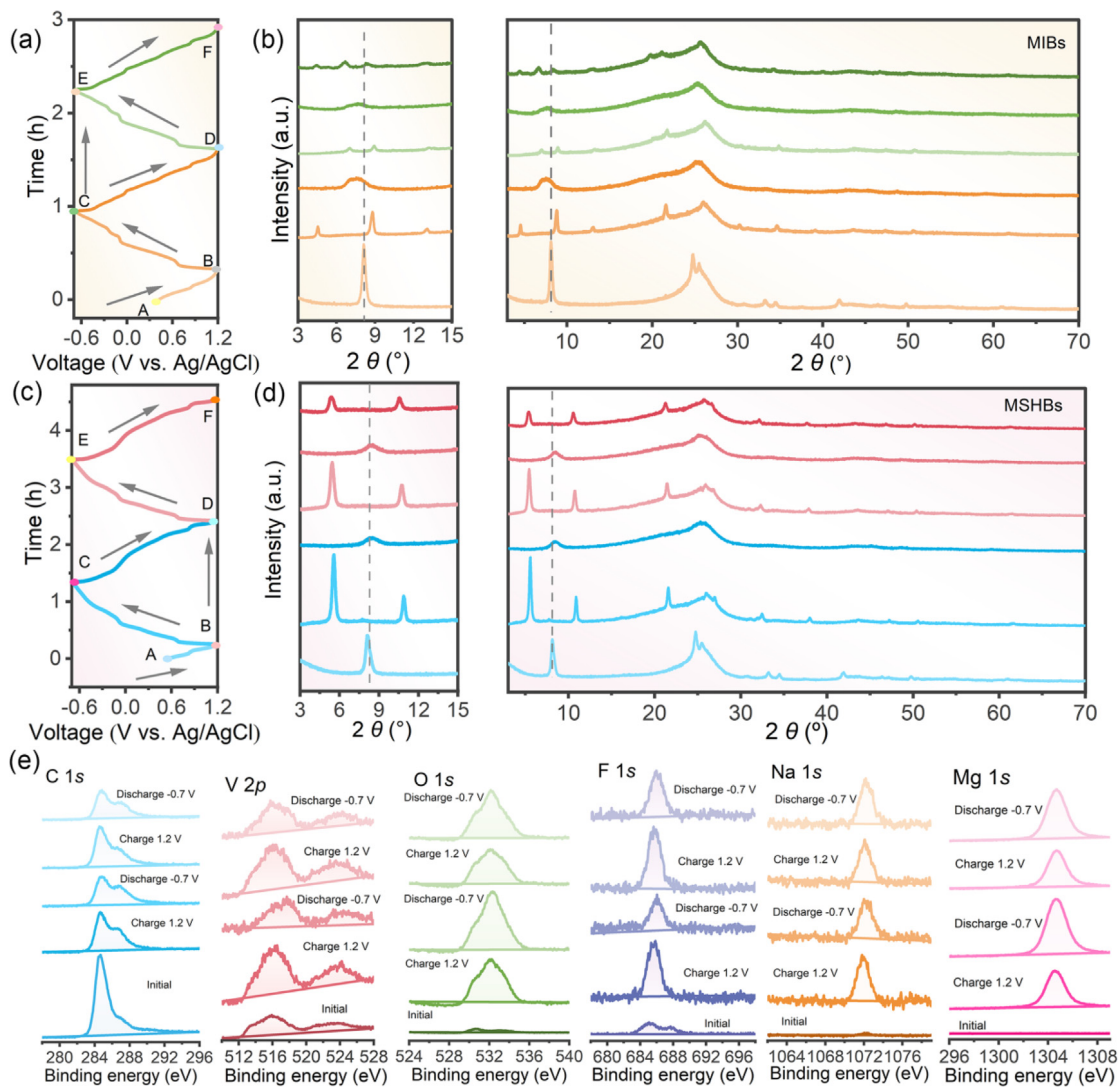

**Fig. 4.** (a) The initial two voltage-time curves of Mn-NVO cathode in MIBs (0.2 A g$^{-1}$); (b) corresponding ex-situ XRD patterns of the Mn-NVO cathode taken from the indicated points in (a); (c) the initial two voltage-time profiles of Mn-NVO cathode in MSHBs (0.2 A g$^{-1}$); (d) corresponding ex-situ XRD patterns of the Mn-NVO cathode taken from the indicated points in (c); (e) XPS depth profiling of C 1$s$, V 2$p$, O 1$s$, F 1$s$, Na 1$s$, and Mg 1$s$ after 140 s for the initial, the first fully charged, the first fully discharged, the second fully charged, the second fully discharged states, and the third fully charged states.

Moreover, ex-situ XRD measurements were conducted at various discharge/charge states to investigate the storage mechanism of MSHBs and SIBs (Fig. S7). The ex-situ XRD pattern in the SIBs exhibits good reversibility (Fig. S8). Fig. S9 displays the ex-situ XRD of cathode in pure aqueous electrolyte (1 M Mg(ClO$_4$)$_2$·6H$_2$O + 1.5 M NaClO$_4$ in H$_2$O), and it is obvious that the structural changes are irreversible during the process of charging and discharging, most likely owing to water damaging the cathode structure. It proves that the added organic solvent TEGDME is also important for stability during cycling. This finding implies that the addition of sodium salts to the MIBs and the retention of Na$^+$ and Mg$^{2+}$ in the interlayer space of Mn-NVO may be critical to the cycling performance, which significantly enhances the capacity retention although a portion of the capacity at low current densities is sacrificed [6]. A portion of Na$^+$ and Mg$^{2+}$ as the pillar can keep the crystal structural stability for the Mn-NVO cathode throughout the cycling process, which can also lead to reduced strain during the Na$^+$ and Mg$^{2+}$ insertion/extraction, so the stability and reversibility are improved [47]. Besides, we consider that the storage mechanism for both Na$^+$ and Mg$^{2+}$ is an intercalation reaction, which may be consistent with the "rocking chair" theory.

Ex-situ XPS investigations were conducted to investigate the ion storage mechanism of the Mn-NVO cathode in MSHBs. On the initial few cycles, Ar$^+$ ion sputtering was also carried out (Fig. 4e) [25]. There is no Mg 1$s$ signal in the initial Mn-NVO cathode. At the fully discharged state, the peak intensity arrives its highest point (discharge to −0.7 V). This observation suggests that Mg$^{2+}$ has been inserted into the Mn-NVO cathode during the discharge process. Moreover, in the fully charged state (charge to 1.2 V), the Mg$^{2+}$ content decreases to a lower value. However, a few of Mg$^{2+}$ that stays within the crystal lattice of the fully charged Mn-NVO cathode acts as pillar ions alongside Na$^+$ to augment the rate capability and cycling stability of the Mn-NVO cathode. The result indicates a reversible process of $Mg^{2+}/Na^+$ insertion and extraction of the Mn-NVO cathode in MSHBs [47].

Besides, the XPS spectra obtained from the Mn-NVO cathodes at various charge and discharge states can also further verify the type of intercalated ions. As demonstrated in Fig. 4(e), when the Mn-NVO cathode is discharged to −0.7 V, there is a noticeable intensification of the peaks corresponding to Na 1$s$ and Mg 2$p$ in the XPS spectra, implying co-insertion of $Mg^{2+}/Na^+$ into the Mn-NVO during the discharge process. Conversely, when the Mn-NVO cathode is charged to 1.2 V, the distinctive peaks associated with Na 1$s$ and Mg 2$p$ are attenuated, revealing that $Mg^{2+}$ and $Na^+$ have been extracted from the Mn-NVO (Table S2). Moreover, in the fully charged state, there is still a minor presence of $Mg^{2+}$ and $Na^+$ within the crystal lattice of the Mn-NVO cathode. This occurrence can be associated to the electrostatic interaction between $Mg^{2+}/Na^+$ and lattice oxygen [10]. By analyzing ex-situ XPS, it is evident that Na and Mg elements still exist in some unrecovered residual chemical states. Because of the consistent and regular change seen in numerous elements during the charging and discharging processes in MSHBs, it is still possible to draw the general conclusion of high reaction reversibility [47].

Fig. 5 presents the findings from HRTEM characterizations and TEM-EDS element mapping, focusing on the distributions of the elements V, O, Na, and Mg. These techniques were employed to investigate the reaction mechanisms of the Mn-NVO cathode within the MSHBs [19]. As displayed in Fig. 5(a–g), the HRTEM of initial Mn-NVO presents great crystallinity of the Mn-NVO cathode. The interplanar spacing of the crystal was found to be 1.08 nm, which corresponds to the (0 0 1) plane of Mn-NVO [10]. Upon reaching the first fully charged state, the interplanar spacing of the crystal changed to 0.83 nm as a result of the deintercalation of $Na^+$ ions, corresponding to the (0 0 2) plane for solvent intercalation of $V_2O_5$. Subsequently, the interplanar distance of the crystal returned to the initial 1.08 nm with the intercalation of $Na^+$ and $Mg^{2+}$ at the first discharged, and maintained a reversible transition during the subsequent cycle process, which confirmed the ex-situ XRD result [25]. The SEM images of different states (initial, the first charge, the first discharge, the second charge, the second discharge, and the third charge states) of Mn-NVO cathodes in MSHBs are shown in Fig. S10. Even and continuous cathode electrolyte interface (CEI) film is present on the cathode surface during cycling, and the addition of TEGDME organic solvent contributes to the stability of the cathode material during cycling [25].

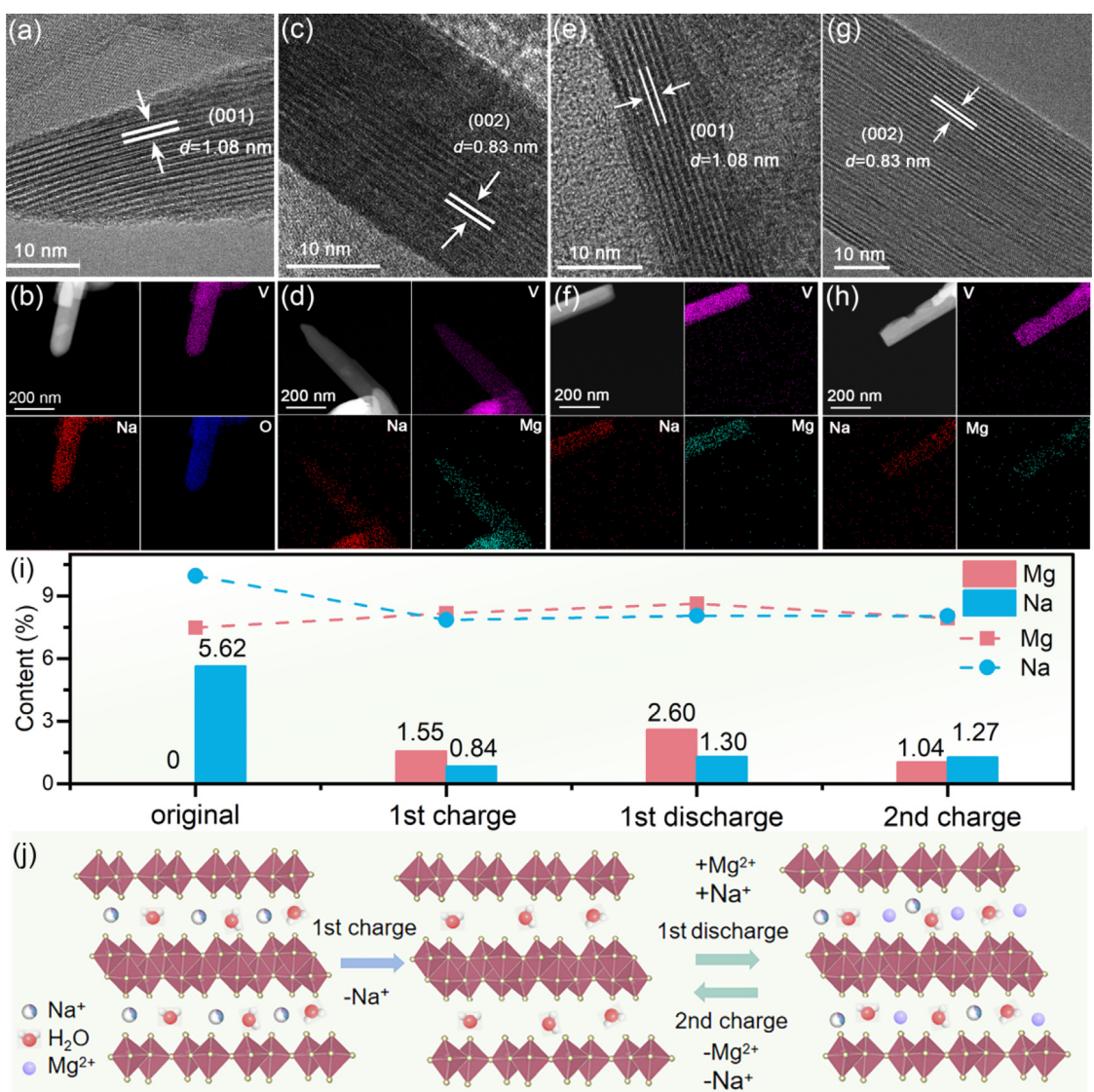

**Fig. 5.** (a–h) HRTEM images of Mn-NVO cathode and corresponding TEM-EDS element mapping for different states (initial, the 1st charged, the 1st discharged, and the 2nd charged state) in MSHBs; (i) TEM-EDS element comparison on Mn-NVO cathode in different states (initial, the 1st charged, the 1st discharged, and the 2nd charged state); (j) schematic diagram illustrating the mechanism of the $Na^+$ extraction from the Mn-NVO cathode, as well as the subsequent $Mg^{2+}/Na^+$ co-insertion charge/discharge process.

TEM-EDS mapping enables the identification of Mg and Na elements within the Mn-NVO cathode in a discharged state. This presents obvious evidence that $Mg^{2+}$/$Na^+$ was inserted into the lattice structure of the cathode material (Fig. 5b–h) [51]. The intensity of the Mg element decreases and increases regularly during charge and discharge in MSHBs, demonstrating the deintercalation and intercalation of $Mg^{2+}$ in the Mn-NVO cathode, respectively [51]. Besides, the intensity of Na decreases and increases weakly during the charge and discharge, which implies that there is a co-deintercalation/intercalation of $Mg^{2+}$ and $Na^+$ in the Mn-NVO cathode. Specifically, as displayed by the TEM-EDS results in MSHBs (Fig. 5i and Table S3), the Mg content from the initial to the second charged state is 0, 1.55, 2.60, and 1.04 at%. Furthermore, the Na contents from the initial to the second charged state are 5.62, 0.84, 1.30, and 1.27 at%. This is obviously not the same as that in MIBs. Besides, the residual $Mg^{2+}$ and $Na^+$ were discovered in a charged state. These ions act as columnar ions, playing a vital role in stabilizing the structure of the cathode material. This enhanced structural stability resulted in great cycling and rate performance [25].

Fig. 5(j) illustrates the reaction mechanism of MSHBs and demonstrates the simultaneous co-intercalation of $Na^+$ and $Mg^{2+}$ ions within the Mn-NVO cathode [33]. The presence of $Na^+$ and $Mg^{2+}$ in the Mn-NVO cathode acts as the structure stabilizing "pillars". These ions maintain the structural stability of the cathode during cycling by reducing the polarization interaction between ions and the cathode. Moreover, this contributes to enhanced cycling stability and rate performance in MSHBs [47].

The electrochemical performance of Mn-NVO cathodes was explored using CR2032 coin cells, where organic PTCDI served as the anode material and 1 M $Mg(ClO_4)_2$ + 1.5 M $NaClO_4$ in TEGDME:$H_2O$ = 4: 1 as electrolytes. To address the challenges posed by slow kinetics and inadequate rate performance in MIBs, leveraging low-cost and abundantly available MSHBs is imperative. The selection of commercial organic PTCDI as the anode material for MSHBs was based on its favorable characteristics, including a low reaction potential (−0.6 V vs. SHE) and theoretical capacity of 137 mA h $g^{-1}$, considering a two-electron transfer per carbonyl group [52].

Besides, the PTCDI anode also exhibits great cycling stability when utilized as a Mg/Na ion storage anode [42]. The working principle of the MSHBs is shown in Fig. 6(a), which reflects that the MSHBs can achieve $Mg^{2+}$/ $Na^+$ co-(de)intercalating into the host materials during the charging and discharging process [8]. XRD pattern of PTCDI anode is demonstrated in Fig. S11. To evaluate the feasibility of PTCDI in MSHBs, the relevant tests in Fig. 6(b and c) demonstrate that it exhibits excellent electrochemical activity in MSHBs [52]. The CV curves of the organic PTCDI anode in MSHBs exhibit a slight peak shift and an increase in peak intensity with increased scan rates, which demonstrates the fast kinetics of organic PTCDI anode in MSHBs (Fig. 6b) [12]. The CV curves of the organic PTCDI anode in SIBs and MIBs are also exhibited in Figs. S12 and S13. Besides, Fig. 6(c) presents the rate performance of the PTCDI anode in the MSHBs. The PTCDI anode in the MSHBs displayed reversible capacities of 101.7 mA h $g^{-1}$ at 0.3 A $g^{-1}$ and 61.1 mA h $g^{-1}$ at 3.0 A $g^{-1}$, showing the reversibility and fast kinetics of ion storage [25]. The CV curves of Mn-NVO//PTCDI full battery kept their shapes from 1.0 to 5.0 mV $s^{-1}$ (Fig. 6d) [42]. Additionally, the CV and charge-discharge curves of Mn-NVO, PTCDI, and the full battery indicate that the assembled full battery can operate within a broad working voltage window spanning from 0 and 1.8 V (Fig. S14) [53]. Fig. 6(e) further reveals the rate performance of the Mn-NVO//PTCDI full cells utilizing Mg/Na hybrid electrolytes at 0.1 and 2 A $g^{-1}$. The reversible capacities achieved at these rates were 97.7 and 57.2 mA h $g^{-1}$, respectively [54,55]. The enhanced capacity can be related to the activation of the electrode material during the first 10 cycles [1,56]. In addition, the electrochemical impedance spectroscopy (EIS) of the Mn-NVO//PTCDI full cell as shown in Fig. S15 can indicate the accelerated kinetics of the MSHBs compared to the MIBs.

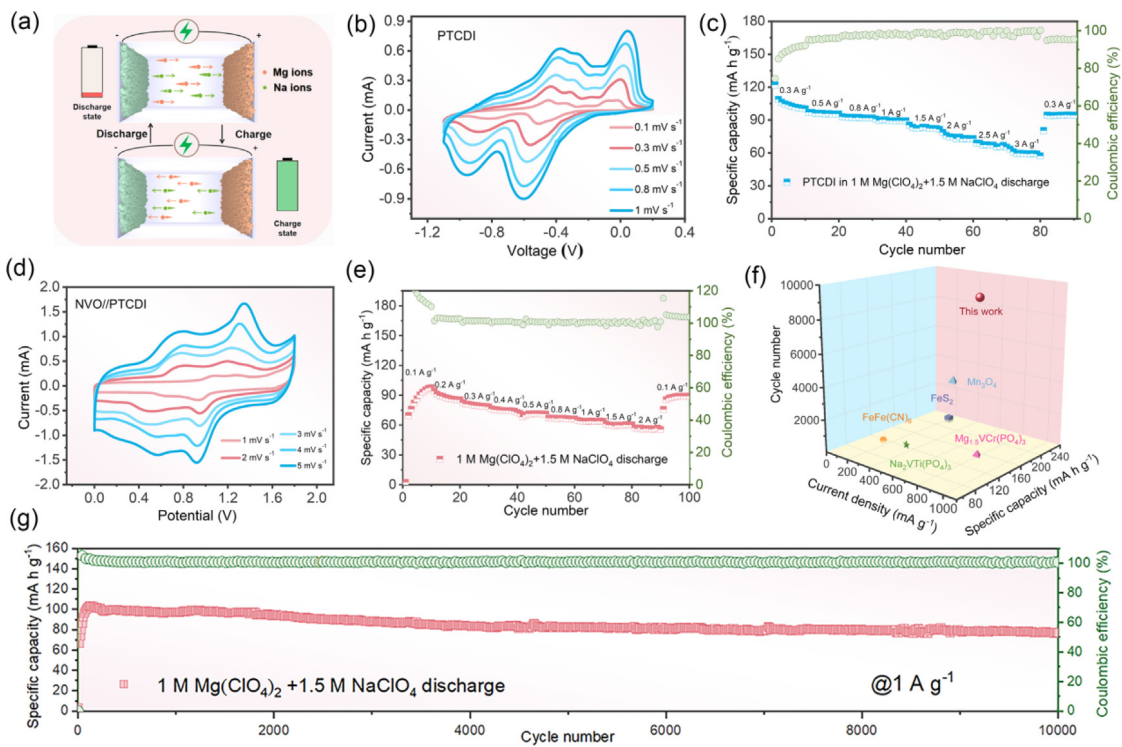

Fig. 6. Comprehensive performance of the full MSHBs. (a) Schematic depiction for the charge and discharge processes of the MSHBs; (b) CV of PTCDI anode (within −1.1–0.2 V) at 0.1, 0.3, 0.5, 0.8, and 1.0 mV $s^{-1}$ in MSHBs; (c) rate performance of PTCDI anode from 0.3 to 3.0 A $g^{-1}$; (d) CV of Mn-NVO//PTCDI full battery within 0–1.75 V in MSHBs at 1.0, 2.0, 3.0, 4.0, and 5.0 mV $s^{-1}$; (e) rate performance of Mn-NVO//PTCDI from 0.1 to 2.0 A $g^{-1}$; (f) electrochemical performance comparison of Mn-NVO//PTCDI with others for MSHBs reported in literatures; (g) the cycling performance of the Mn-NVO//PTCDI full battery at 1 A $g^{-1}$.

The Mn-NVO//PTCDI full cells indicate outstanding electrochemical performance, surpassing the results reported in previous research on MSHBs (i.e. FeS$_2$//Mg [21], Na$_2$VTi(PO$_4$)$_3$//Mg [22], Mn$_3$O$_4$//NaTi$_2$(PO$_4$)$_3$ [23], Mg$_{1.5}$VCr(PO$_4$)$_3$//FeVO$_4$ [20], and FeFe(CN)$_6$//Mg [24]) (Fig. 6f and Table S1) [18]. In addition, NVO//PTCDI full cell also preserved the discharge capacity of 77.4 mA h g$^{-1}$ after 10,000 cycles at 1 A g$^{-1}$, confirming great cycling stability of the full cell (Fig. 6g) [4]. In the first few cycles, the low coulombic efficiency and discharge capacity can be ascribed to the difficulty of ion intercalation. When the electrode material is activated, the occurrence of side reactions between the electrode material and the electrolyte diminishes, resulting in an enhancement in the chemical reaction as the active sites become more exposed [1]. Fig. S16 illustrates the charge-discharge voltage curves of Mn-NVO//PTCDI full MSHBs at several cycle numbers, demonstrating excellent cycle stability. Furthermore, we conducted a comparative analysis of the energy density of the full battery with recent reported works, as shown in Fig. S17 and Table S4. Remarkably, the cells in this study exhibit a higher energy density (97.7 Wh kg$^{-1}$) in comparison to the other magnesium-based cells. This outcome underscores the exceptional electrochemical performance of MSHBs, revealing a promising design approach for rechargeable batteries in practical energy storage applications [57–59].

## 4. Conclusions

Briefly, a novel MSHB has been developed, incorporating Mn-NVO as the cathode and PTCDI as the anode, paired with a hybrid electrolyte consisting of 1.5 M NaClO$_4$ + 1.0 M Mg(ClO$_4$)$_2$ in TEGDME: H$_2$O = 4: 1. The environmental friendliness and earth-rich composition are clear advantages of MSHBs. Significantly, a synergistic effect emerges between the Mn-NVO cathode and hybrid electrolyte, playing a crucial role in enabling rapid and stable storage of Mg$^{2+}$ and Na$^+$ ions. Simultaneously, the Mn-NVO cathode presents a great discharge-specific capacity (249.9 mA h g$^{-1}$ at 0.3 A g$^{-1}$), high-rate capability (130.2 mA h g$^{-1}$ at 3 A g$^{-1}$), and remarkable cycling stability in MSHBs (1500 cycles at 1.5 A g$^{-1}$). Furthermore, the Mn-NVO//PTCDI full battery has an exceptional cyclic stability of 10,000 cycles at 1.0 A g$^{-1}$. Overall, we expect that this work will demonstrate the potential of MSHBs with excellent electrochemical properties and cost-effective characteristics to enhance the cycling stability and rate capability of conventional MIBs and boost the progress of aqueous hybrid batteries.

## CRediT authorship contribution statement

**Xiaoke Wang:** Data curation, Methodology, Writing – original draft. **Titi Li:** Data curation, Investigation. **Xixi Zhang:** Formal analysis, Validation. **Yaxin Wang:** Data curation, Methodology. **Hongfei Li:** Data curation, Supervision, Writing – review & editing. **Hai-Feng Li:** Methodology, Supervision, Writing – review & editing. **Gang Zhao:** Methodology, Supervision, Validation, Writing – review & editing. **Cuiping Han:** Conceptualization, Investigation, Methodology, Resources, Supervision, Writing – review & editing.

## Declaration of competing interest

The authors declare that they have no known competing financial interests or personal relationships that could have appeared to influence the work reported in this paper.

## Acknowledgments

We acknowledge the financial support from the National Natural Science Foundation of China, China (22005207, 52261160384), and the Guangdong Basic and Applied Basic Research Foundation, Guangdong Province, China (2019A1515011819), the Outstanding Youth Basic Research Project of Shenzhen, Shenzhen, China (RCYX20221008092934093), the Joint Funds of the National Natural Science Foundation of China, China (U22A20140), and the Science and Technology Development Fund, Macao SAR (0090/2021/A2 and 0049/2021/AGJ).

## Appendix A. Supplementary material

Supplementary data to this article can be found online at https://doi.org/10.1016/j.jechem.2024.04.016.

## References

[1] X. Cheng, Z. Zhang, Q. Kong, Q. Zhang, T. Wang, S. Dong, L. Gu, X. Wang, J. Ma, P. Han, H.-J. Lin, C.-T. Chen, G. Cui, Angew. Chem. Int. Ed. Engl. 59 (2020) 11477–11482.
[2] G. Liu, Y. Tang, H. Li, J. He, M. Ye, Y. Zhang, Z. Wen, X. Liu, C.C. Li, Angew. Chem. Int. Ed. Engl. 62 (2023) e202217945.
[3] S. Ding, X. Dai, Z. Li, C. Wang, A. Meng, L. Wang, G. Li, J. Huang, S. Li, Energy Storage Mater. 47 (2022) 211–222.
[4] Y. Xu, W. Cao, Y. Yin, J. Sheng, Q. An, Q. Wei, W. Yang, L. Mai, Nano Energy 55 (2019) 526–533.
[5] W. Chen, X. Zhan, R. Yuan, S. Pidaparthy, A.X.B. Yong, H. An, Z. Tang, K. Yin, A. Patra, H. Jeong, C. Zhang, K. Ta, Z.W. Riedel, R.M. Stephens, D.P. Shoemaker, H. Yang, A.A. Gewirth, P.V. Braun, E. Ertekin, J.-M. Zuo, Q. Chen, Nat. Mater. 22 (2023) 92–99.
[6] Z. Yang, X.-H. Liu, X.-X. He, W.-H. Lai, L. Li, Y. Qiao, S.-L. Chou, M. Wu, Adv. Funct. Mater. 31 (2021) 2006457.
[7] Y. Zhang, Z. Yuan, L. Zhao, Y. Li, X. Qin, J. Li, W. Han, L. Wang, Small (2023) 2301815.
[8] X. Yu, G. Zhao, C. Liu, C. Wu, H. Huang, J. He, N. Zhang, Adv. Funct. Mater. 31 (2021) 2103214.
[9] C. Pérez-Vicente, S. Rubio, R. Ruiz, W. Zuo, Z. Liang, Y. Yang, G.F. Ortiz, Small 19 (2023) 2206010.
[10] D. Wu, F. Wang, H. Yang, Y. Xu, Y. Zhuang, J. Zeng, Y. Yang, J. Zhao, Energy Storage Mater. 52 (2022) 94–103.
[11] S. Rubio, R. Liu, X. Liu, P. Lavela, J.L. Tirado, Q. Li, Z. Liang, G.F. Ortiz, Y. Yang, J. Mater. Chem. A 7 (2019) 18081–18091.
[12] Z. Chen, Q. Yang, D. Wang, A. Chen, X. Li, Z. Huang, G. Liang, Y. Wang, C. Zhi, ACS Nano 16 (2022) 5349–5357.
[13] C.-H. Shin, H.-Y. Lee, C. Gyan-Barimah, J.-H. Yu, J.-S. Yu, Chem. Soc. Rev. 52 (2023) 2145–2192.
[14] Y. Xu, Z. Liu, X. Zheng, K. Li, M. Wang, W. Yu, H. Hu, W. Chen, Adv. Energy Mater. 12 (2022) 2103352.
[15] Y. Liu, W. Zhao, Z. Pan, Z. Fan, M. Zhang, X. Zhao, J. Chen, X. Yang, Angew. Chem. Int. Ed. Engl. 62 (2023) e202302617.
[16] S. Zhang, Y. Wang, Y. Sun, Y. Wang, Y. Yang, P. Zhang, X. Lv, J. Wang, H. Zhu, Y. NuLi, Small 19 (2023) 2300148.
[17] Y. Han, G. Li, Z. Hu, F. Wang, J. Chu, L. Huang, T. Shi, H. Zhan, Z. Song, Energy Storage Mater. 46 (2022) 300–312.
[18] H. Xu, X. Zhang, T. Xie, Z. Li, F. Sun, N. Zhang, H. Chen, Y. Zhu, X. Zou, C. Lu, J. Zou, R.M. Laine, Energy Storage Mater. 46 (2022) 583–593.
[19] Y. Ding, T. Han, Z. Wu, Y. Guan, J. Hu, C. Hu, Y. Tian, J. Liu, ACS Nano 16 (2022) 15369–15381.
[20] Y. Tang, X. Li, H. Lv, W. Wang, Q. Yang, C. Zhi, H. Li, Angew. Chem. Int. Ed. Engl. 60 (2021) 5443–5452.
[21] M. Walter, K.V. Kravchyk, M. Ibáñez, M.V. Kovalenko, Chem. Mater. 27 (2015) 7452–7458.
[22] Y. Zhang, J. Gui, T. Li, Z. Chen, S.-A. Cao, F. Xu, Chem. Eng. J. 399 (2020) 125689.
[23] X. Cao, L. Wang, J. Chen, J. Zheng, J. Mater. Chem. A 6 (2018) 15762–15770.
[24] H. Dong, Y. Li, Y. Liang, G. Li, C.-J. Sun, Y. Ren, Y. Lu, Y. Yao, Chem. Commun. 56 (2016) 8263–8266.
[25] X. Wang, X. Zhang, G. Zhao, H. Hong, Z. Tang, X. Xu, H. Li, C. Zhi, C. Han, ACS Nano 16 (2022) 6093–6102.
[26] S. Li, J. Shang, M. Li, M. Xu, F. Zeng, H. Yin, Y. Tang, C. Han, H.-M. Cheng, Adv. Mater. 35 (2022) 2207115.
[27] S. Hou, X. Ji, K. Gaskell, P.-F. Wang, L. Wang, J. Xu, R. Sun, O. Borodin, C. Wang, Science 374 (2021) 172–178.
[28] P. Wang, J. Trück, J. Häcker, A. Schlosser, K. Küster, U. Starke, L. Reinders, M.R. Buchmeiser, Energy Storage Mater. 49 (2022) 509–517.
[29] M. Karlsmo, R. Bouchal, P. Johansson, Angew. Chem. Int. Ed. Engl. 60 (2021) 24709–24715.
[30] G. Zhu, G. Xia, H. Pan, X. Yu, Adv. Sci. 9 (2022) 2106107.

[31] X. Zhao, J. Yan, H. Hong, Y. Zhao, Q. Li, Y. Tang, J. He, Z. Wei, S. He, X. Hou, C. Zhi, H. Li, Adv. Energy Mater. 12 (2022) 2202478.
[32] X. Xu, F. Xiong, J. Meng, X. Wang, C. Niu, Q. An, L. Mai, Adv. Funct. Mater. 30 (2020) 1904398.
[33] C. Wang, L. Zhang, M. Al-Mamun, Y. Dou, P. Liu, D. Su, G. Wang, S. Zhang, D. Wang, H. Zhao, Adv. Energy Mater. 9 (2019) 1900909.
[34] M. Du, C. Liu, F. Zhang, W. Dong, X. Zhang, Y. Sang, J.-J. Wang, Y.-G. Guo, H. Liu, S. Wang, Adv. Energy Mater. 7 (2020) 2000083.
[35] Y. Dong, L. Miao, G. Ma, S. Di, Y. Wang, L. Wang, J. Xu, N. Zhang, Chem. Sci. 12 (2021) 5843–5852.
[36] G. Ma, L. Miao, Y. Dong, W. Yuan, X. Nie, S. Di, Y. Wang, L. Wang, N. Zhang, Energy Storage Mater. 47 (2022) 203–210.
[37] C. Yang, Z. Pu, Z. Jiang, X. Gao, K. Wang, S. Wang, Y. Chai, Q. Li, X. Wu, Y. Xiao, D. Xu, Adv. Energy Mater. 12 (2022) 2201718.
[38] Y. Cao, Y. Zhu, C. Du, X. Yang, T. Xia, X. Ma, C. Cao, ACS Nano 16 (2022) 1578–1588.
[39] Y. Yang, G. Qu, H. Wei, Z. Wei, C. Liu, Y. Lin, X. Li, C. Han, C. Zhi, H. Li, Adv. Energy Mater. 13 (2023) 2203729.
[40] Y. Zhu, J. Yin, X. Zheng, A.-H. Emwas, Y. Lei, O.F. Mohammed, Y. Cui, H.N. Alshareef, Energy Environ. Sci. 14 (2021) 4463–4473.
[41] X. Li, Q. Li, Y. Hou, Q. Yang, Z. Chen, Z. Huang, G. Liang, Y. Zhao, L. Ma, M. Li, Q. Huang, C. Zhi, ACS Nano 15 (2021) 14631–14642.
[42] G. Liang, Z. Gan, X. Wang, X. Jin, B. Xiong, X. Zhang, S. Chen, Y. Wang, H. He, C. Zhi, ACS Nano 15 (2021) 17717–17728.
[43] S. Zheng, D. Shi, D. Yan, Q. Wang, T. Sun, T. Ma, L. Li, D. He, Z. Tao, J. Chen, Angew. Chem. Int. Ed. Engl. 61 (2022) e202117511.
[44] X. Lin, J. Liu, H. Zhang, Y. Zhong, M. Zhu, T. Zhou, X. Qiao, H. Zhang, T. Han, J. Li, Adv. Energy Mater. 8 (2021) 2002298.
[45] X. Wang, H. Dong, A. Eddine Lakraychi, Y. Zhang, X. Yang, H. Zheng, X. Han, X. Shan, C. He, Y. Yao, Mater. Today 55 (2022) 29–36.
[46] H. Chen, H. Qin, L. Chen, J. Wu, Z. Yang, J. Alloys Compd. 842 (2020) 155912.
[47] S. Ding, X. Dai, Y. Tian, B. Song, L. Wang, G. Li, S. Li, J. Huang, Z. Li, A. Meng, Nano Energy 93 (2022) 106838.
[48] X. Qu, A. Du, T. Wang, Q. Kong, G. Chen, Z. Zhang, J. Zhao, X. Liu, X. Zhou, S. Dong, G. Cui, Angew. Chem. Int. Ed. Engl. 61 (2022) e202204423.
[49] R. Zhang, C. Cui, R. Xiao, L. Ruinan, T. Mu, H. Huo, Y. Ma, G. Yin, P. Zuo, Chem. Eng. J. 444 (2022) 136592.
[50] X.-F. Ma, H.-Y. Li, X. Zhu, W. Ren, X. Zhang, J. Diao, B. Xie, G. Huang, J. Wang, F. Pan, Small 18 (2022) 2202250.
[51] K.W. Leong, W. Pan, Y. Wang, S. Luo, X. Zhao, D.Y.C. Leung, ACS Energy Lett. 7 (2022) 2657–2666.
[52] R. Li, J. Yu, F. Chen, Y. Su, K. Chan, Z. Xu, Adv. Funct. Mater. 33 (2023) 2214304.
[53] E. Li, M. Wang, Y. Feng, L. Yang, Q. Li, Z. Yang, J. Chen, B. Yu, B. Guo, Z. Ma, Y. Huang, J. Liu, X. Li, J. Energy Chem. 94 (2024) 148–157.
[54] L. Ma, X. Li, G. Zhang, Z. Huang, C. Han, H. Li, Z. Tang, C. Zhi, Energy Storage Mater. 31 (2020) 451–458.
[55] Y. Jing, Q. Lv, Y. Chen, B. Wang, B. Wu, C. Li, S. Yang, Z. He, D. Wang, H. Liu, S. Dou, J. Energy Chem. 94 (2024) 158–168.
[56] X. Yu, G. Zhao, C. Wu, H. Huang, C. Liu, X. Shen, M. Wang, X. Bai, N. Zhang, J. Mater. Chem. A 9 (2021) 23276–23285.
[57] J. Bae, H. Park, X. Guo, X. Zhang, J.H. Warner, G. Yu, Energy Environ. Sci. 14 (2021) 4391–4399.
[58] J. Wang, L. Jiao, Q. Liu, W. Xin, Y. Lei, T. Zhang, L. Yang, D. Shu, S. Yang, K. Li, C. Li, C. Yi, H. Bai, Y. Ma, H. Li, W. Zhang, B. Cheng, J. Energy Chem. 94 (2024) 10–18.
[59] G. Zhao, Y. Xing, Y. Liu, X. Wang, B. Zhang, L. Mu, W. Liao, X. Xu, Mater. Today Chem. 34 (2023) 101758.